\documentclass[reprint,amssymb, amsmath, aip,cha,twocolumn]{revtex4-1}
\usepackage{graphicx}
\usepackage[caption = false]{subfig}
\usepackage{color}
\usepackage{epsfig}
\usepackage{ifpdf}
\usepackage{url}%
\usepackage{multirow}
\usepackage{float}
\usepackage{bm}
\usepackage[colorlinks=true,linkcolor=blue]{hyperref}%
\usepackage{cleveref}
\expandafter\ifx\csname package@font\endcsname\relax\else
 \expandafter\expandafter
 \expandafter\usepackage
 \expandafter\expandafter
 \expandafter{\csname package@font\endcsname}% 
\fi
\begin{document}
\title{Reflection of a dust acoustic solitary wave in a dusty plasma}%
\author{Krishan Kumar}%
\email{kumar.krishan861@gmail.com}
\author{P. Bandyopadhyay}
\author{Swarnima Singh}
\author{Garima Arora}
\author{A. Sen}
\affiliation{$^1$Institute For Plasma Research, HBNI, Bhat, Gandhinagar, Gujarat, India, 382428}%
\date{\today}
%**************************************************************
%#####################################################################################                  ABSTRACT
%************************************************************************************************
\begin{abstract} 
We report the first experimental observations of the reflection of a dust acoustic solitary wave from a potential barrier in a dusty plasma medium. The experiments have been carried out in an inverted $\Pi$-shaped Dusty Plasma Experimental (DPEx) device in a DC glow discharge plasma environment. The dust acoustic solitary wave is excited by modulating the plasma with a short negative Gaussian pulse that is superimposed over the discharge voltage. The solitary wave structure is seen to move towards a potential barrier, created by the sheath around a biased wire, and turn back after reflecting off the barrier. The  amplitude, width, and velocity of the soliton are recorded as a function of  time. The experiment is repeated for different strengths of the potential barrier and for different initial amplitudes of the solitary wave. It is found that the distance of the closest approach of the solitary wave to the centre of the barrier increases with the increase of the strength of the potential barrier and with the decrease of the initial wave amplitude. An emissive probe is used to measure the sheath potential and its thickness by measuring the plasma potential profile in the axial direction over a range of resistances connected to the biased wire. A modified Korteweg de Vries equation is derived and numerically solved to qualitatively understand the experimental findings.
\end{abstract}
%%%%%%%%%%%%%%%%%%%%%%%%%%%%%%%%
\maketitle
%%%%%%%%%%%%%%%%%%%%%%%%%%%%%%%%%%%%%%%%%%%%%%%%%%%%%%%%%%%%%%%%%%%%%%%%%%%%%%%%%%%%%       INTRODUCTION
%%%%%%%%%%%%%%%%%%%%%%%%%%%%%%%%%%%%%%%%%%%%%%%%%%%%%%%%%%
 %***********************************************************************************************************************
 \section{Introduction}\label{sec:intro}
 Solitons are self-reinforcing wave packets that can propagate over long distances, preserving their shapes and velocities, and are remarkable nonlinear structures that were first observed in water waves and subsequently found in many different media. Mathematical models describing their evolution show that their unique property arises from an exact balance between non-linear wave steepening and dispersive wave broadening\cite{drazin_johnson_1989}. Colliding solitons also exhibit ``elastic'' behavior - a ``particle'' like property that has given rise to their nomenclature \cite{drazin_johnson_1989, Sharma_2014}.  Solitons and solitary waves are ubiquitous in nature and are also seen in many laboratory experiments. They have been identified in space plasma fluctuations \cite{Trines_2007}, ocean waves \cite{New_1990}, propagating pulses in optical fibres \cite{Gedalin_1997,Haus_1996}, semiconductor plasmas \cite{Barland_2002}, Bose-Einstein condensates \cite{Bludov_2009}, and a host of other plasma and fluid systems \cite{Zabusky_1965,Heidemann_2009,Samsonov_2002,Bandyopadhyay_2008,Washimi_1966,Nakamura_2001}. Several model  non-linear evolution equations that are fully integrable yield soliton solutions. The Korteweg-de Vries (KdV) equation \cite{Zabusky_1971,Sen_2015} is one such non-linear partial differential equation that has been extensively employed as a model to study low-frequency non-linear wave phenomena in plasmas under conditions of weak dispersion and weak non-linearity.\par 
The reflection and transmission of a solitonic pulse in an inhomogeneous medium is a topic that has attracted considerable interest in the past e.g.  in Bose-Einstein condensates \cite{Marchant_2013} and in electron-ion plasmas \cite{Dahiya_1978,Nishida_1984,Nagasawa_1986,Cooney_1991, Raychaudhuri_1986, Kuehl_1983, Chauhan_1996, Malik_2014}. In a plasma solitons can be reflected due to the presence of a density gradient or from externally imposed barriers such as a metal plate. Dahiya \textit{et al.}\cite{Dahiya_1978} experimentally showed the partial reflection of ion-acoustic solitary waves by a negatively biased grid in the plasma. Nishida \textit{et al.}\cite{Nishida_1984} demonstrated the reflection of a planar ion-acoustic soliton from a finite plane boundary. Nagasawa \textit{et al.} \cite{Nagasawa_1986} investigated the reflection and the refraction of ion-acoustic solitons from a metallic mesh electrode situated in a double-plasma device. Cooney \textit{et al.} \cite{Cooney_1991} experimentally investigated the propagation of a soliton, its collision and reflection near the sheath boundary in a multicomponent plasma and discussed the conservation law of the reflection and transmission of a soliton. Raychaudhuri \textit{et al.} \cite{Raychaudhuri_1986} showed experimentally the reflection of ion-acoustic solitons from a bipolar potential structure and explained the mechanism of the reflection. Kuehl \textit{et al.} \cite{Kuehl_1983} studied theoretically the reflection of an ion-acoustic soliton and found that the amplitude of the reflected soliton is much smaller than that of the incident soliton. Chauhan \textit{et al.} \cite{Chauhan_1996} investigated the propagation and the reflection of an ion-acoustic soliton in an inhomogeneous plasma. Malik \textit{et al.}\cite{Malik_2014} derived the reflection and transmission coefficients of an ion-acoustic soliton.\par
A large number of experimental\cite{Samsonov_2002,Bandyopadhyay_2008,Jaiswal_2016}, theoretical\cite{Shukla_2003,Rao_1998,27Shukla_2003,Rao_2001,Bharuthram_1992} and simulation\cite{Sen_2015,Popel_2003,Tiwari_2016} studies have been carried out to study the propagation characteristics of Dust Acoustic Solitary Waves (DASWs) and solitons in a dusty plasma. A dusty (complex) plasma consists of micron or sub-micron-sized particles, electrons, ions, and neutrals. These micron-sized particles  usually get negatively charged in the plasma background  by collecting more electrons than ions and levitate near the plasma sheath boundary. These massive and highly charged dust particles immersed in a conventional plasma introduce novel linear and non-linear collective modes with very low frequency. The reflection of dust acoustic solitary waves in a dusty plasma medium remains as yet an unexplored area of research. In this paper, we  investigate the reflection of a dust acoustic solitary wave off an external potential barrier in a dusty plasma. The dust acoustic solitary wave is excited by modulating the plasma with a short negative Gaussian pulse that is superimposed over the discharge voltage.  The wave first propagates in the forward direction toward the potential barrier and reaches a threshold point (point of reflection) and then it reflects. The distance of the closest approach is found to be directly dependent on the strength of the potential barrier and the initial soliton amplitude. To understand these experimental findings, a modified KdV equation is constructed using the full set of fluid equations in the presence of an external Gaussian potential. The numerical solution of this KdV equation qualitatively reproduces results that are seen in the experiments. 
%We believe the first-ever experimental observation of the reflection of dust acoustic solitary waves in a dusty plasma will surely encourage the researchers to contribute more research work in this direction. \par 
The paper is organized as follows: in Sec.~\ref{sec:setup} the experimental apparatus and details of the experimental procedure are described. Sec.~\ref{sec:results} contains the experimental findings and a brief discussion on them. A theoretical model to describe the experimental results is presented in Sec.~\ref{sec:model}. Sec.~\ref{sec: summary} provides a summary and some concluding remarks.

\section{Experimental apparatus and procedure}\label{sec:setup}
\begin{figure}[ht]
\includegraphics[scale=0.35]{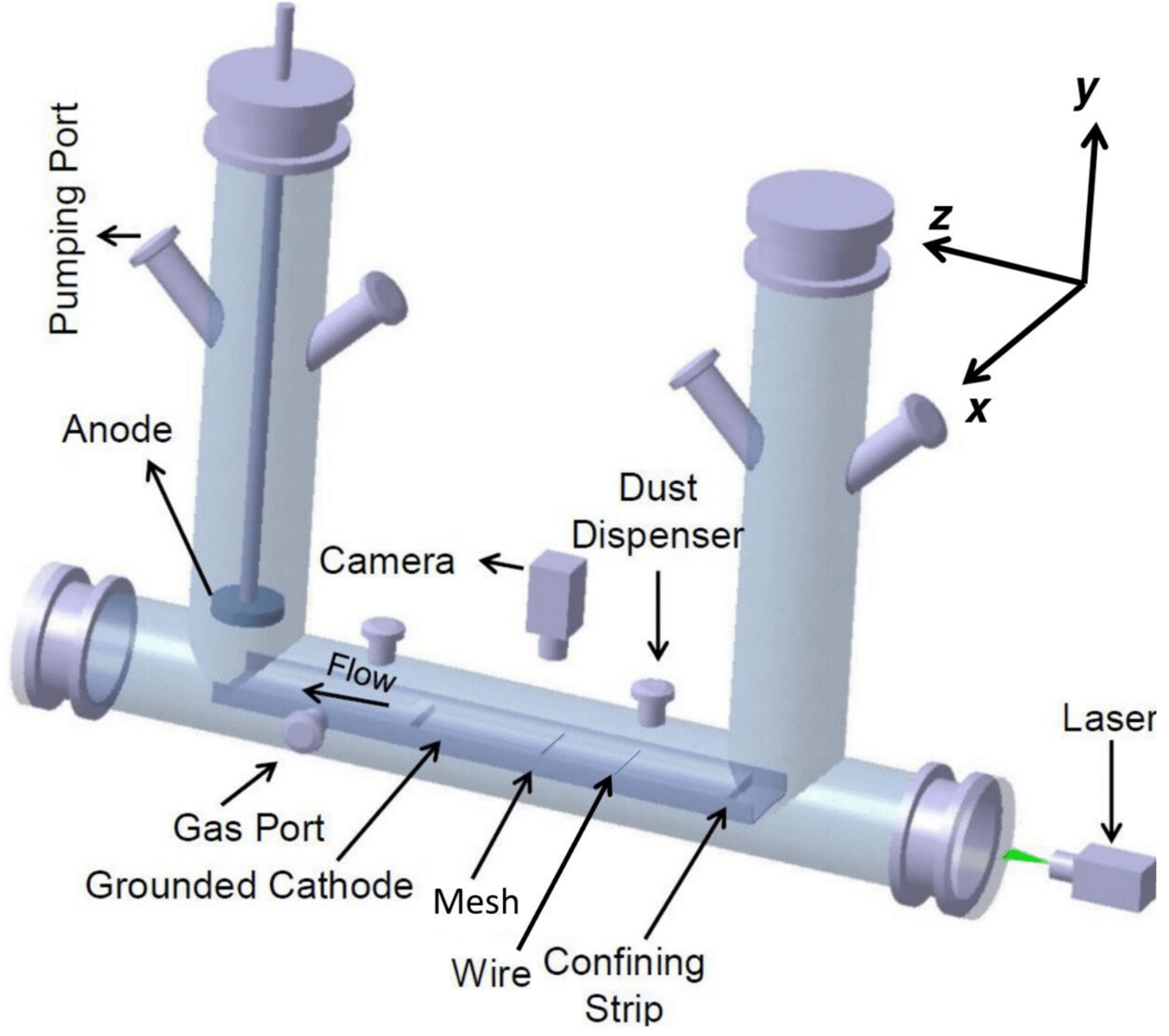}
\caption{\label{fig:fig1} A schematic diagram of dusty plasma experimental (DPEx) device.  }
\end{figure}

\begin{figure}[ht]
\includegraphics[scale=0.75]{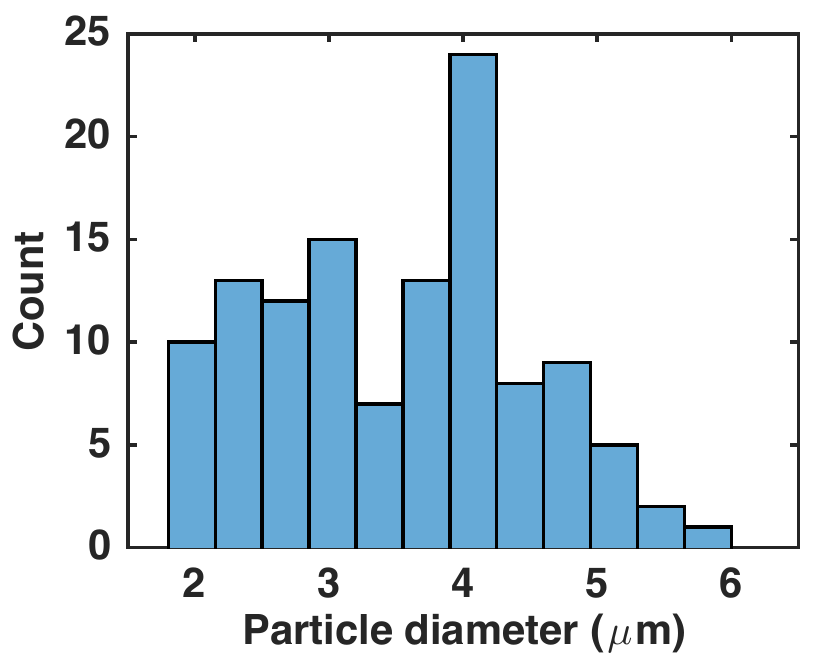}
\caption{\label{fig:distribution} The size distribution of the Kaolin particles which are used as dust component in the experiment.}
\end{figure}

The present set of experiments are carried out in an inverted $\Pi$-shaped Dusty Plasma Experimental (DPEx) device, whose schematic diagram is shown in Fig.~\ref{fig:fig1}. The details of the experimental device and its related diagnostics are discussed in Ref.~\cite{Jaiswal_2015}. 
%%%%%%%%%%%%%%%%%%% FIGURE   %%%%%%%%%%%%%%%%%%%%%
%%%%%%%%%%%%%%%%%%%%%%%%%%%%%%%%%%%%%%%%%%%%%%%%%%%%
To evacuate the vacuum vessel to its base pressure of 0.1~Pa, a rotary pump is attached in one of the auxiliary tubes, while a couple of Pirani gauges are installed at two different places to monitor the gas pressure in the chamber. A circular SS plate is used as an anode and a tray-shaped long SS grounded plate is used as a cathode. The bent edges of the cathode provide the radial confinement to the particles whereas, for axial confinement, two metallic rectangular strips are used. One of them is kept exactly at the edge of the cathode, whereas another one is placed at a distance of approximately 30~cm. To excite the solitary wave in the present set of experiments, a grounded fine mesh is installed at a distance of 20~cm from the cathode edge, and additionally, a copper wire of diameter of 1.0~mm is mounted horizontally at a height of 1 cm from the cathode plate at a distance of 13~cm from the right end of the cathode. The sheath around this wire serves as the reflector for the dust acoustic solitary wave. There is  provision to keep this wire either in grounded potential or in floating potential. The wire can also be maintained at an intermediate potential with respect to the grounded potential by drawing plasma current through a variable resistance connected in series with the wire and the ground \cite{Arora_2019}. Poly-dispersive Kaolin particles are used as dust component, which are initially filtrated through a fine mesh having sieve separation of $\sim 8$~$\mu$m. Scanning Electron Microscopic (SEM) images of these Kaolin particles are used to obtain their size distribution. A typical size distribution of Kaolin particles is shown in Fig.~\ref{fig:distribution}. Fig.~\ref{fig:distribution} essentially shows that the filtrated Kaolin particles have a random size distribution of $\sim$~2--5~$\mu$m. These poly-dispersive micron sized Kaolin particles are then heated up to remove the water content and are sprinkled on the cathode in between wire and mesh before closing the chamber.
 %%%%%%%%%%%%%%%%%%% FIGURE   %%%%%%%%%%%%%%%%%%%%%
%%%%%%%%%%%%%%%%%%%%%%%%%%%%%%%%%%%%%%%%%%%%%%%%%%%%
Such poly-dispersive kaolin particles have been extensively used in the past to experimentally study the propagation characteristics of linear and nonlinear waves in dusty plasmas \cite{Barkan_1995, Pramanik_2002, Bandyopadhyay_2008, Heinrich_2009,Jaiswal_2016}. Likewise, many past experiments have also used mono-dispersive particles to successfully study the excitations of various non-linear waves in a dusty plasma medium \cite{Samsonov_2002,Pieper_1996,Heidemann_2009,Samsonov_2004}. \par
To begin with, the chamber is pumped down with the help of the rotary pump to attain a base pressure of 0.1 Pa. Argon gas is then fed into the chamber using a mass flow controller to set the working pressure to 8-10 Pa. A DC glow discharge Argon plasma is formed in between the anode and the grounded tray cathode by applying a voltage in the range of 280-360~V using a DC power supply. Plasma parameters like plasma density ($n_{i}$) $\sim$ 0.5--3 $ \times$ $10^{15} $ $~m^{-3} $ and electron temperature ($ T_{e} $) $\sim$ 2--5~eV are  measured using a single Langmuir probe over a  range of discharge parameters \cite{Jaiswal_2015}. In the plasma environment, the dust particles lying on the cathode get negatively charged and levitate at the cathode sheath in between the wire and the mesh where the electrostatic force acting on the negatively charged particles balances the gravitational force. Due to the size distribution of the Kaolin particles as shown in Fig.~\ref{fig:distribution}, the heavier (bigger) particles levitate at the bottom, whereas the lighter (smaller) particles levitate at the top in $y-z$ plane. A green line laser of width $\sim 0.5$~mm is used to illuminate the levitated dust particles of the central region of the cloud, where the dust density is maximum. In that particular layer, the particles are assumed to be spherical of average radius 1.75~$\mu$m for the estimation of their average charge and mass. A CCD camera is used for capturing the dynamics of dust particles in a computer. The recorded sequences of images are then analysed with the help of MATLAB.
\par
For the present set of experiments, the discharge voltage and the neutral gas pressure are maintained at 350~V and 8.6~Pa, respectively.  For this particular discharge condition,the  plasma and dusty plasma parameters such as  plasma density, $n_{i} \sim 1\times 10^{14}$ $~m^{-3} $, electron temperature $ T_{e} \sim$ 4~eV, dust density  $ n_{d} \sim 5 \times 10^{9}$ $~m^{-3} $ and the average dust mass $ m_{d} $ $ \sim $ 6.0 $ \times $ $ 10^{-14} $~kg are used to estimate the other dusty plasma parameters. The average charge on a dust particle $ Q_{d} $ is estimated to be $ \sim 2.5 \times 10^{4}$e. To get an idea of the coupling between the charged dust particles, the screened Coulomb coupling parameter is estimated for a typical discharge condition. The Coulomb coupling parameter is given by the expression \cite{Melzer_1996,Thomas_1994,Merlino_2012},
$ \Gamma=\frac{Q_{d}^{2}}{4\pi\epsilon_{0}ak_{B}T_{d}}exp(-\frac{a}{\lambda_{D}})$,
where, $T_d, a$, and $\lambda_D$ are the dust temperature, inter-particle distance, and dust Debye length, respectively. $\epsilon_0$ is the permittivity of free space, and $k_B$ is the Boltzmann constant. For the present set of experiments, these parameters come out to be $a \sim 360~\mu m$, $ \lambda_{D}\sim 117~\mu m $. The average dust temperature is estimated as $ T_{d} \sim 12.40$~eV by tracking the individual particles of the  dust cloud for 100 consecutive highly resolved camera images using a super particle identification tracking (sPIT) code \cite{Feng_2007}. The relatively high kinetic energy of the dust particles can be attributed to the presence of ambient electric field fluctuations in the plasma, as has been noted and reported  many times in the past  \cite{Merlino_2012, Heinrich_2009}. With the help of these parameters, the coupling strength ($ \Gamma $) of the medium is estimated to be $\sim 9$, which essentially indicates that the dust component is in the fluid regime. Its fluid nature is also evident  from the camera images of the dust cloud that resemble the images of the experiments carried out in the past by Heinrich \textit{et al.} \cite{Merlino_2012, Heinrich_2009}. The acoustic velocity in the dusty plasma medium is estimated from the expression\cite{Merlino_2012}, $C_d =Z_d\sqrt{\frac{n_dKT_i}{m_dn_i}}$, which yields a value of $4.6$~cm/s. Before exciting the non-linear dust acoustic wave (DAW), a controlled experiment is performed to trigger a linear dust acoustic wave by applying a sinusoidal electrical pulse to the mesh with a frequency and amplitude of 1Hz and 30~V, respectively.  It is found that the dust acoustic wave propagates through the dusty plasma medium with a velocity of 5.0-6.0 ~cm/sec, which is found to be very close to the estimated phase velocity of a linear DAW. Later, the plasma is modulated to excite nonlinear waves in the dusty plasma medium by following the same technique used in the past \cite{Bandyopadhyay_2008} and will be discussed in the following section. 
%%%%%%%%%%%%%%%%%%%%%%%%%%%%%%%%%%
\section{Experimental Results}\label{sec:results}
\subsection{Excitation and characterization of solitary wave}
\begin{figure}[h]
\hspace*{-0.4 cm}
\includegraphics[scale=0.70]{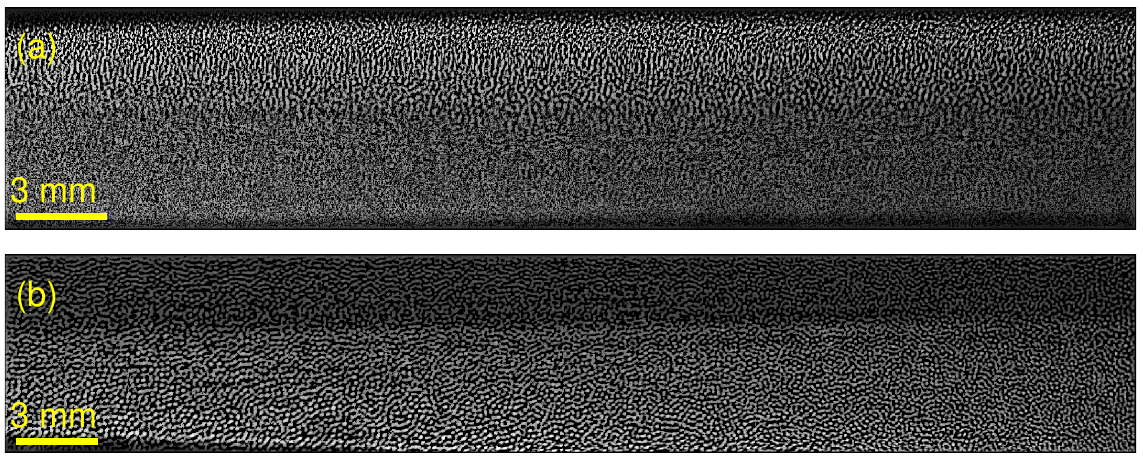}
\caption{\label{fig:equilibrium} (a) Horizontal (x-z) and (b) vertical (y-z) views of equilibrium dust cloud.}
\end{figure}

\begin{figure}[h]
\hspace*{-0.4 cm}
\includegraphics[scale=1.0]{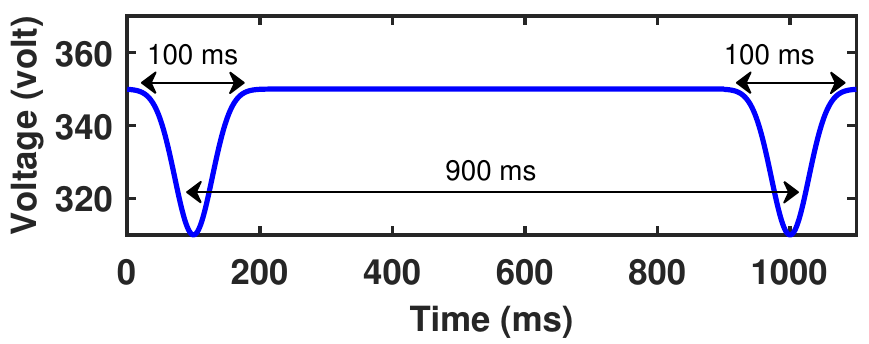}
\caption{\label{fig:fig2} Typical voltage pulse used to excite the dust acoustic solitary wave.}
\end{figure}

\begin{figure}[h]
\includegraphics[scale=0.88]{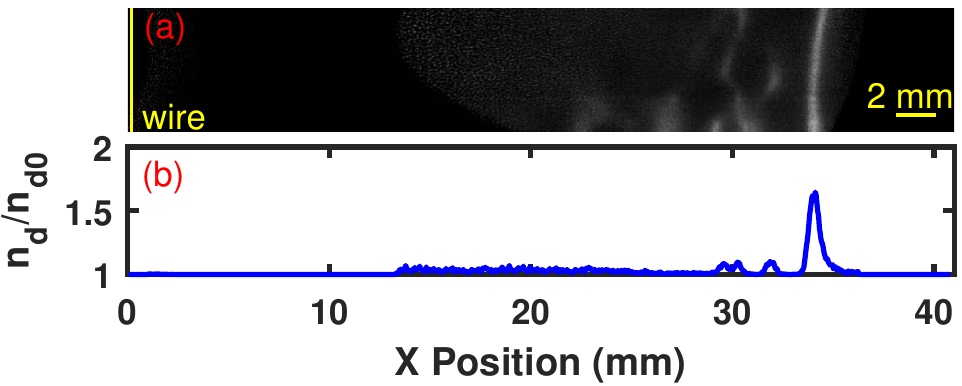}
\caption{\label{fig:fig3} (a) A snapshot of the excited nonlinear wave in the dusty plasma medium. (b) The intensity profile of the solitary wave.}
\end{figure}

Before the excitation of a dust acoustic solitary wave, the equilibrium dust cloud is examined to check whether there are any  ion streaming instability\cite{Merlino_2009} induced spontaneous excitation of Dust Acoustic Waves (DAW)  in the medium at the working pressure of 8.6~Pa. Fig.~\ref{fig:equilibrium}(a) and (b) show the horizontal (x-z) and vertical (y-z) views of the equilibrium dust cloud respectively.  As can be seen the equilibrium dust density images show no signs of any disturbances indicating the absence of any spontaneous excitation of  DAW at this working pressure.
%%%%%%%%%%%%%%%%%%% FIGURE   %%%%%%%%%%%%%%%%%%%%%
%%%%%%%%%%%%%%%%%%%%%%%%%%%%%%% 
For the excitation of a non-linear dust acoustic solitary wave in the dusty plasma medium, the plasma is modulated and this plasma modulation technique is employed from the earlier work of Bandyopadhyay \textit{et al.} \cite{Bandyopadhyay_2008}. A negative Gaussian pulse is superimposed on the discharge voltage to perturb the equilibrium dust cloud. The on-time and the off-time of the applied pulse are set to be 100~ms and 900~ms respectively as shown in fig.~\ref{fig:fig2}. The amplitude of the pulse is chosen in such a way that it leads to the excitation of a single dust acoustic solitary wave. 
 %%%%%%%%%%%%%%%%%%% FIGURE   %%%%%%%%%%%%%%%%%%%%%
%%%%%%%%%%%%%%%%%%%%%%%%%%%%%%% 
The negative pulse creates a small density perturbation in the dust cloud and the compressive dust density perturbation begins to move in the forward direction.  Fig.~\ref{fig:fig3}(a) shows a typical snapshot of the excitation of dust acoustic solitary wave at the time when the plasma is modulated with a Gaussian pulse. During this initial stage of excitation of the solitary wave, comparatively smaller amplitude disturbances occur in the medium, which later disappear. The density compression of the excited dust acoustic solitary wave is shown in fig.~\ref{fig:fig3}(b). It can be seen clearly from fig.~\ref{fig:fig3}(a) that the single prominent compressive structure gets excited in the dusty plasma medium in between the wire and the mesh. The sharp peak and high amplitude as shown in fig.~\ref{fig:fig3}(b) essentially indicate that the structure is highly non-linear in nature. It is also to be noted that the small peaks in the intensity profiles, as shown in Fig.~\ref{fig:fig3}(b), indicate the small disturbances along with the well-defined wave structure. \par
%%%%%%%%%%%%%%%%%%% FIGURE   %%%%%%%%%%%%%%%%%%%%%
\begin{figure}[h]
\includegraphics[scale=0.85]{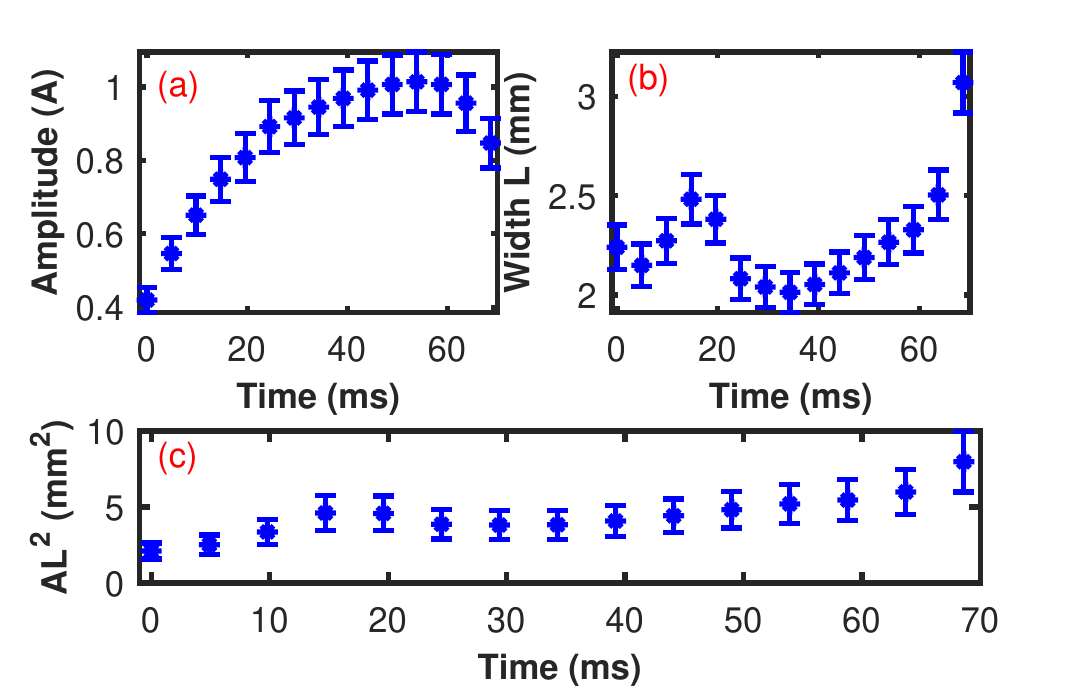}
\caption{\label{fig:fig4} Variation of (a) solitary amplitude, (b) solitary width and (c) the product of solitary amplitude and the square of its width with time. }
\end{figure}

%%%%%%%%%%%%%%%%%%% FIGURE   %%%%%%%%%%%%%%%%%%%%% 
To characterise this excited wave, the amplitude, $A$ (defined as $n_{d}/n_{d0}-1 $, where $ n_{d}$ and $n_{d0}$ are the instantaneous and equilibrium dust densities, respectively) and width, $L$ (full width at half maxima) of the non-linear  wave are measured over  time and plotted in fig.~\ref{fig:fig4}(a) and fig.~\ref{fig:fig4}(b), respectively. The amplitude of the solitary wave, first increases, attains a constant value, and then decreases. However, the width initially rises and then follows an opposite trend to that of the amplitude. It may be due to the fact the wave initially grows with time and then propagates with almost constant amplitude and width and then finally it decays. Interestingly, the product of the amplitude ($A$) and the square of the width ($L$) as calculated over time is found to remain constant during its journey as shown in fig~\ref{fig:fig4}(c), similar to that of  KdV type solitons \cite{Bandyopadhyay_2008,Miura_1976}.
%%%%%%%%%%%%%%%%%%% FIGURE   %%%%%%%%%%%%%%%%%%%%%
%%%%%%%%%%%%%%%%%%% FIGURE   %%%%%%%%%%%%%%%%%%%%%
%%%%%%%%%%%%%%%%%%% FIGURE   %%%%%%%%%%%%%%%%%%%%%
\begin{figure}[h]
\includegraphics[scale=1.0]{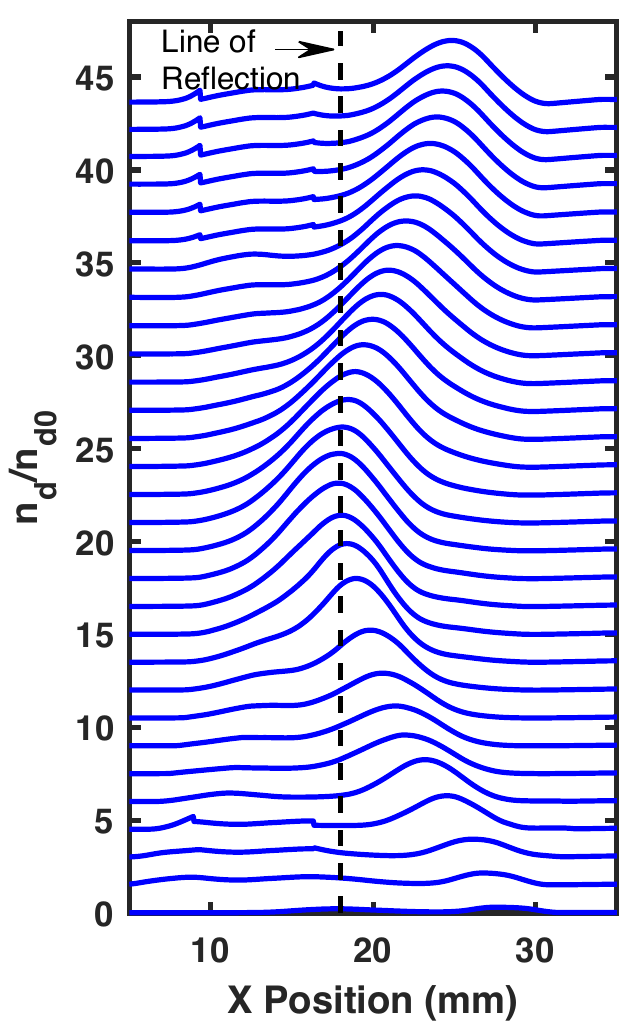}
\caption{\label{fig:fig5} Time evolution of the dust acoustic solitary wave. The compression factor ($n_d/{n_{d0}}$) is plotted in the interval of 4.9~ms. The dashed line represents the line of reflection. The wire (not shown in figure) is kept at \lq 0' location, whereas the mesh is located at $X=70$~mm.}
\end{figure}  
%%%%%%%%%%%%%%%%%%%%%%%%%%%%%%%%%
\subsection{Interaction of solitary wave with sheath potential}
\begin{figure}[h]
\includegraphics[scale=0.6]{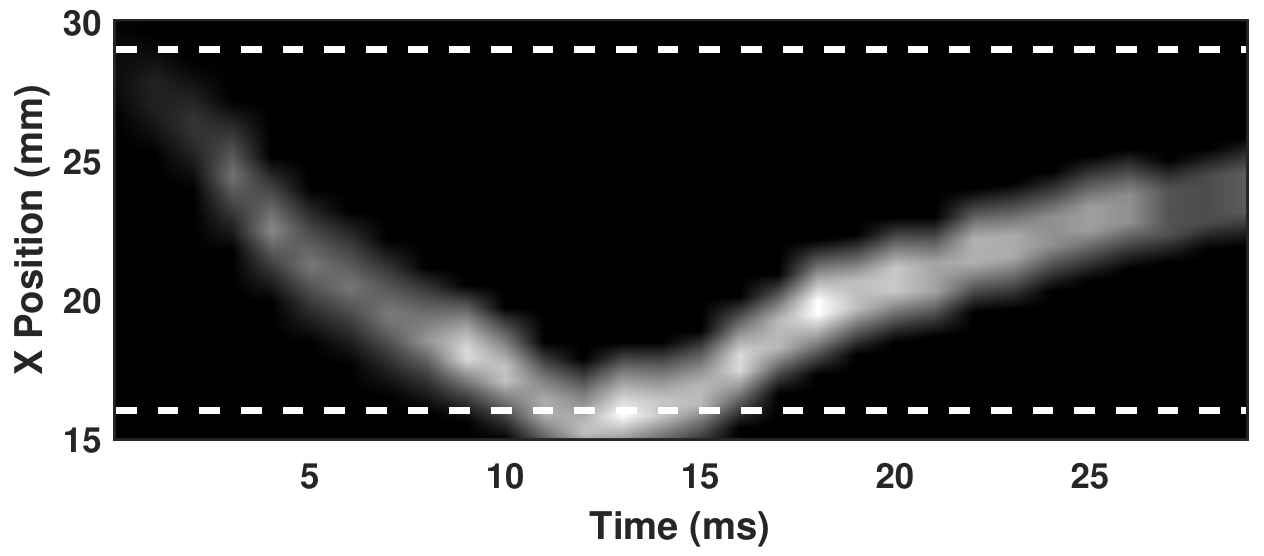}
\caption{\label{fig:period}The periodogram plot of the dust acoustic solitary wave.}
\end{figure}

To examine the characteristics of this solitary wave while propagating towards a biased wire several experiments have been carried out by varying the resistance of the potentiometer connected with the wire and the ground. Fig.~\ref{fig:fig5} shows the typical time evolution of the solitary wave in the case when the wire is kept at floating potential. It is clearly seen in the figure, that the solitary wave gets excited with smaller amplitude at a distance of $\sim$~28~mm away from the copper wire, then grows up and propagates toward the wire. The solitary wave feels the effect of the sheath potential as it comes closer to the wire. The kinetic energy of the solitary wave keeps on decreasing as it enters into the pre-sheath region and propagates further towards the wire. At the point of reflection, the wave kinetic energy becomes exactly equal to the potential energy which makes the wave stationary. After reaching this point the solitary wave propagates in the opposite direction. In this specific case, after reaching a closest approach distance of 17~mm, the solitary wave turns around and starts to propagate in the opposite direction. It is to be noted that, after travelling a distance of $\sim$ 9 mm in the backward direction the solitary wave gets dissipated in the medium due to dust-neutral collisions (not shown in fig.~\ref{fig:fig5}). To study the reflection of the dust acoustic solitary wave in more details, a periodogram\cite{Schwabe_2007} showing  its complete evolution is displayed in Fig.~\ref{fig:period}. The periodogram, is actually a space-time diagram, which provides a visual indication of the characteristics of a propagating wave. It is clearly seen in Fig~\ref{fig:period}, that the solitary wave gets excited with smaller amplitude and then it grows up with time. After propagating a distance the solitary wave gains a constant amplitude and reaches to the reflection point.  Afterwards it reflects back and propagates in the opposite direction almost with the same amplitude. Finally, after travelling a certain distance its amplitude decays as discussed in fig.~\ref{fig:fig5}. \par 
%%%%%%%%%%%%%%%%%%% FIGURE   %%%%%%%%%%%%%%%%%%%%%
\begin{figure}[h]
\includegraphics[scale=0.3]{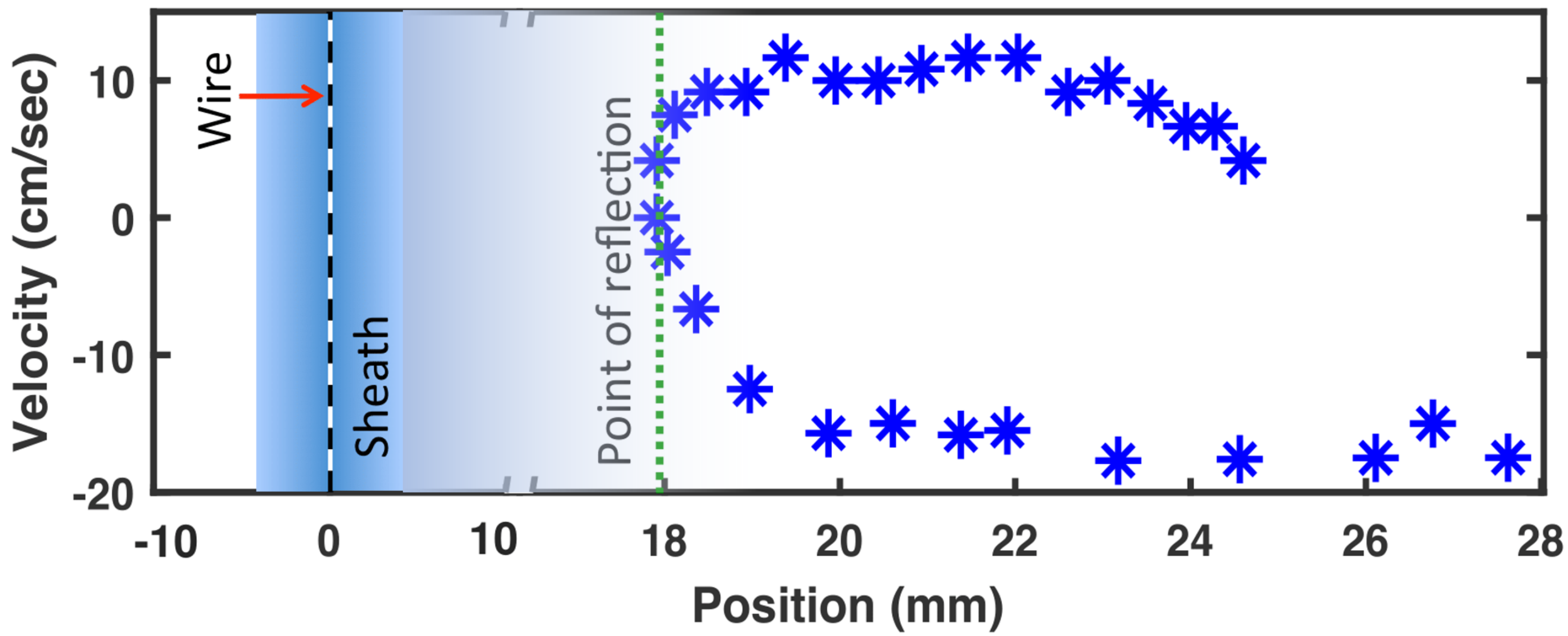}
\caption{\label{fig:fig6} Phase-space diagram of the solitary wave. Dashed and dotted lines represent the position of the wire and the point of reflection, respectively. The mesh is located at a position of 70 mm. The shaded region represents the sheath region around the wire.}
\end{figure}

%%%%%%%%%%%%%%%%%%%%%%%%%%%%%%%%%
Fig.~\ref{fig:fig6} depicts the phase-space diagram of dust acoustic solitary wave in the case when the wire is biased at floating potential. The vertical dashed line represents the position of the wire whereas the shaded region in Fig.~\ref{fig:fig6} indicates the sheath around the wire. It is clear from this figure that the solitary wave propagates initially with a constant velocity of $\sim$~15~cm/sec and then slows down as it approaches towards the negative sheath and turns back at the reflection point (indicated by dotted line). After reflection, the solitary wave first accelerates, and then it recovers almost its original velocity of $\sim$~12~cm/sec. It is also worth mentioning that after reaching the closest approach of 17~mm, the solitary wave becomes almost stationary and starts to propagate in the opposite direction as shown also in Fig.~\ref{fig:fig5}. \par
%%%%%%%%%%%%%%%%%%% FIGURE   %%%%%%%%%%%%%%%%%%%%%
\begin{figure}[h]
\includegraphics[scale=1.0]{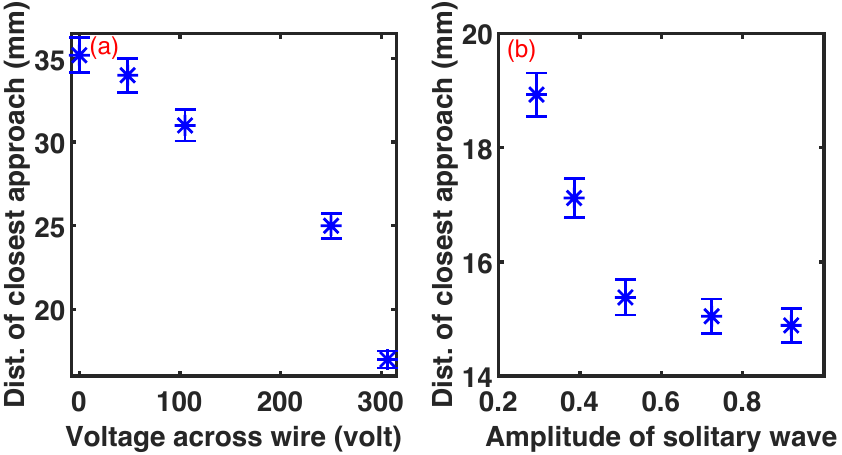}
\caption{\label{fig:fig7} Variation of the distance of the closest approach of the solitary wave from the wire with (a) voltage across the wire and (b) amplitude of the solitary wave. }
\end{figure}

%%%%%%%%%%%%%%%%%%%%%%%%%%%%%%%%%
To investigate the interaction of dust acoustic solitary wave with the electrostatic reflector (negative sheath around the wire) in a detailed manner, the experiments are repeatedly performed by altering the negative sheath potential around the wire. The sheath potential is varied by drawing current through the wire using an external potentiometer of resistance ranging from 100~K$\Omega$ to 10~M$ \Omega$ and measuring the voltage across it. Fig.~\ref{fig:fig7}(a) shows the distance of the closest approach of the solitary wave from the wire for different potential strengths. The distance of the closest approach is defined as the distance between the point of reflection and the wire. The increase in the voltage across the wire results in the decrease of the strength of the potential barrier as well as the sheath thickness \cite{Arora_2019} and hence the distance of the closest approach becomes larger for the higher strength of the potential barrier. It is also worth mentioning that the distance of the closest approach is maximum when the wire is kept at grounded potential whereas it becomes minimum when the wire is biased at floating potential. Earlier studies of Bandyopadhyay \textit{et al.} \cite{Bandyopadhyay_2008} suggest that the soliton amplitude, as well as its velocity, increases with the increase of the discharge pulse height. To investigate the effect of the amplitude of the excited solitary waves on the distance of the closest approach, the Gaussian pulse height (as shown in Fig.~\ref{fig:fig2}) is changed in a controlled manner.  Fig.~\ref{fig:fig7}(b) shows the variation of the distance of the closest approach of the solitary wave with the initial amplitude of the solitary wave. It is observed that the distance of the closest approach decreases with the increase in the amplitude of the solitary wave. The velocity of the solitary wave increases with the increase of its amplitude, which allows the wave to penetrate the sheath even deeper before it gets reflected.\par
 %%%%%%%%%%%%%%%%%%% FIGURE   %%%%%%%%%%%%%%%%%%%%%
%%%%%%%%%%%%%%%%%%%%%%%%%%%%%%%%%
 \subsection{Measurements of sheath thickness and potential}
 \label{sheath}
\begin{figure}[h]
\includegraphics[scale=1]{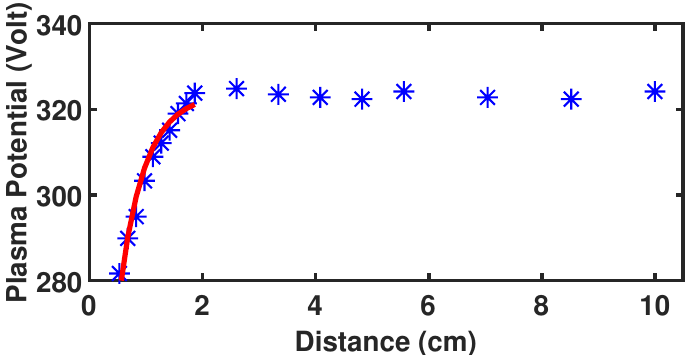}
\caption{\label{fig:fig8} Typical plasma potential profile in axial direction from the wire. }
\end{figure}

 An emissive probe is used to measure the plasma potential to infer directly the sheath thickness and the sheath potential around the wire over a range of biased voltages. A hairpin-shaped tungsten wire of diameter 0.125~mm and length 1~cm with a ceramic holder is used as an emissive probe in our experiments. The plasma potential is measured using the floating point technique as discussed in ref.~\cite{Sheehan_2011}. In this technique, the emissive probe measures the plasma potential by emitting the thermionic electrons which nullifies the ion sheath beneath the probe and allows the probe to measure the plasma potential even in the sheath region. The emissive probe is scanned axially to measure the plasma potentials for different strengths of wire voltage. Initially, the wire is biased at grounded potential and the probe is scanned gradually towards the wire from a distance of 10~cm. Fig.~\ref{fig:fig8} depicts the variation of the plasma potential with the axial probe position, where \lq0' indicates the location of the wire. It clearly shows that the plasma potential remains almost constant in the bulk plasma and falls sharply near the wire due to the presence of ion sheath around it. Past measurements of the sheath potential around a charged object have established that the shape of the potential is nearly Gaussian\cite{Arora_2018}. Therefore, the sheath thickness is estimated from this axial profile (along $z$) of the plasma potential which can be approximately fitted to a Gaussian shape. The fitted curve is also shown in Fig.~\ref{fig:fig8} as a solid line
 superposed on the experimental points.
The sheath thickness is estimated as the half width at full maximum of the Gaussian form 
and yields a value of  $\sim$~1.7~cm for the case of the grounded wire.
%%%%%%%%%%%%%%%%%%% FIGURE   %%%%%%%%%%%%%%%%%%%%%
%%%%%%%%%%%%%%%%%%%%%%%%%%%%%%%%%
The same exercise to evaluate the sheath thickness is followed by increasing the biased voltage of the wire through increasing the resistance of the potentiometer connected in series with the wire and the grounded potential. \par
\begin{figure}[h]
%\hspace*{1. cm}
\includegraphics[scale=1.0]{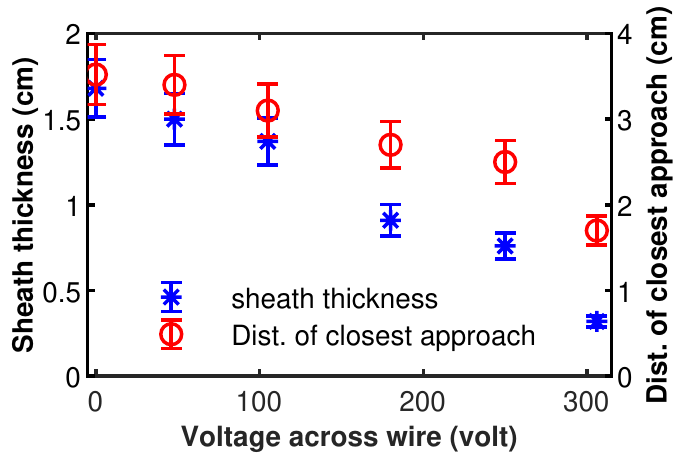}
\caption{\label{fig:fig9} Variation of sheath thickness and the distance of the closest approach with the voltage across the wire.}
\end{figure}

Fig.~\ref{fig:fig9} shows the variation of sheath thickness and the distance of the closest approach with the biased voltage across the wire. One can see that as the voltage across the wire increases the sheath thickness decreases. The sheath thickness depends on the total potential drop with respect to the plasma potential for a given discharge condition. As the voltage across the wire increases, the potential drop decreases, hence the sheath thickness also decreases. Sheath thickness becomes maximum when the wire is at grounded potential and minimum when it is at floating potential. Therefore, the solitary wave propagates more when the wire is at the floating potential and hence the distance of the closest approach becomes minimum as compared to the case when the wire is at the grounded potential. 
%which is also shown in fig.~\ref{fig:fig7}(a).\par
%%%%%%%%%%%%%%%%
%%%%%%%%%%%%%%%%%%%%%%    
\section{Theoretical Model}\label{sec:model} 
To provide theoretical support to our experimental results, we have developed a model equation in the form of  a modified-KdV equation from the basic fluid equations describing a dusty plasma.  In such a system the very low frequency dust-acoustic waves satisfy $\omega<<kv_{Te}, kv_{Ti}$ (where $k$ is the wave number, $v_{Te}=\sqrt{\frac{T_{e}}{m_{e}}} $, and $v_{Ti}=\sqrt{\frac{T_{i}}{m_{i}}}  $  are the ion and electron thermal speeds, $T_{e}~(T_{i})$ and $m_{e}~(m_{i})$ are the temperature and mass of the electron (ion), respectively). Compared to the massive dust particles the electrons and ions can be treated as inertia less plasma species such that their number densities can be described by Boltzmann distributions at temperature $T_e$ and $T_i$ respectively, namely,
%%%%%%%%%%%%%%
\begin{equation}
 n_{e}=n_{e0}\text{exp}\left(\frac{e\phi}{T_{e}}\right),~ n_{i}=n_{i0}\text{exp}\left(\frac{-e\phi}{T_{i}}\right).
\label{eq:Boltzmann distributions}
\end{equation}            
%%%%%%%%%%%%%%
To study the dust dynamics we use the following set of fluid equations,
%%%%%%%%%%%%%%%%%%%%%%%%%%%%%%%%%%%
\begin{eqnarray}
\frac{\partial n_{d}}{\partial t}&+&\frac{\partial (n_{d}v_{d})} {\partial x}=0,
\label{eq:continuity}\\
\frac{\partial v_{d}}{\partial t}&+&v_{d}\frac{\partial v_{d}}{\partial x}-\frac{Z_{d} e}{m_{d}}\frac{\partial \phi}{\partial x}=0,
\label{eq:momentum}\\
\frac{\partial^{2} \phi}{\partial x^{2}}&+&4\pi e(n_{i}-n_{e}-Z_{d}n_{d})=0.
\label{eq:Possion}
\end{eqnarray}
%%%%%%%%%%%%%%%%%%%%%%%%%%%%%%%%%%%
where, $n_{d},~v_{d},~Z_{d},~m_{d}~ \text{and} ~\phi $ are the density, velocity, charge, mass of the dust particles and electrostatic wave potential, respectively. It is to be noted that the dissipative effect due to dust-neutral collision is not considered in the momentum equation (in Eq.~\ref{eq:momentum}). Dust neutral collision frequency for our experiments can be estimated from the expression, $\nu_{dn}=\frac{4}{3}\delta \pi a^2m_nn_nC_n/m_d$ \cite{Epstein_1924}, where $ m_{n}, n_{n}, C_{n}$ are the mass, number density, average velocity of neutral particles, respectively. $\delta$ is the Epstein drag coefficient, which has been experimentally determined for our device  to be $\sim 1.2$ \cite{Jaiswal_2015}. For Argon gas, $ m_{n}\sim 6.63\times 10^{-26}~kg $, $ n_{n}\sim 2\times 10^{21}~m^{-3} $ at a pressure of 8.6~Pa and $ C_{n}\sim 429~m/s$. For these experimental parameters, the value of $\nu_{dn}$ is $\sim$ 15~sec$^{-1}$. As reported in the past, the energy of a soliton decays as $e^{-\nu_{dn}t}$ whereas its width and amplitude change as $e^{\nu_{dn}t/3}$  and $e^{-2\nu_{dn}t/3}$, respectively \cite{Samsonov_2002}. The damping length \cite{Samsonov_2002} for our experiment is therefore approximately estimated to be $3v_s/2\nu_{dn}\sim$ 16~mm for an average soliton velocity ($v_s$) value of  $\sim$ 15~cm/s. For our experiments, it is found that the damping length is eight times larger than the width ($\Delta \sim $2~mm) of the soliton. Therefore, it can be assumed that the dust-neutral collision is not so important in our experiment and hence the dissipative term due to dust neutral collisions is neglected in the theoretical model. To obtain the KdV equation, we use the reductive perturbation analysis technique and expand the variables density, velocity and electrostatic potential by a smallness parameter $\epsilon$ as follows,
%%%%%%%%%%%%%%%%%%%%%%%%%%%%%%%%%%%
\begin{equation}
\Psi=\Psi_{0}+\epsilon\Psi_{1}+\epsilon^{2}\Psi_{2}+\epsilon^{3}\Psi_{3}.
\label{eq:Expansion}
\end{equation}
%%%%%%%%%%%%%%%%%%%%%%%%%%%%%%%%%%%
where $\Psi=(n_d,~v_d,~\phi)$. A point of departure from the usual homogeneous plasma treatment is that the equilibrium quantities,  denoted with a subscript \lq0',  are now weakly varying functions of the spacial coordinate, $x$. This is to account for the weak spatial variation of these quantities near the tail of the sheath region around the wire as the soliton approaches the barrier. Note that due to the existence of a finite $\phi_0(x)$ there will be a corresponding spatially varying $n_{d0}(x)$ and a drift velocity $v_{d0}(x)$ of the dust. These three quantities, more specifically their gradients are related to each other in a self-consistent manner as we will show below. For a dusty plasma medium with spatially varying quantities, we take a suitable set of stretched coordinates introduced by Singh and Rao \textit{et al.}\cite{Singh_1998}:
%%%%%%%%%%%%%%%%%%%%%%%%%%%%%%%%%%%
\begin{equation}
\xi=\epsilon^{1/2}\left( \int^{x}\frac{dx^{'}}{\lambda(x^{'})}-t \right)  \text{and } ~ \eta=\epsilon^{3/2} x.
\label{eq:stretched coordinates}
\end{equation}
%%%%%%%%%%%%%%%%%%%%%%%%%%%%%%%%%%%
where $\epsilon$ is a smallness parameter and $\lambda$ is the velocity of the moving (wave) frame. Since we are considering only spatial gradients, hence,
%%%%%%%%%%%%%%%%%%%%%%%%%%%%%%%%%%%
\begin{equation}
\frac{\partial n_{d0}}{\partial \xi}=0;~\frac{\partial \lambda}{\partial \xi}=0;~\frac{\partial \phi_{0}}{\partial \xi}=0;~\frac{\partial v_{d0}}{\partial \xi}=0
\label{eq:spatial_condition}
\end{equation}
%%%%%%%%%%%%%%%%%%%%%%%%%%%%%%%%%%%
Using Eqs.~\ref{eq:stretched coordinates} and~\ref{eq:spatial_condition} into Eqs.~\ref{eq:continuity}~--~\ref{eq:Possion}, we can express the continuity, momentum and Poisson equations of the dust fluid in the following form, 
%%%%%%%%%%%%%%%%%%%%%%%%%%%%%%%%%%%
\begin{eqnarray}
&-&\lambda \frac{\partial n_{d}}{\partial \xi}+\frac{\partial (n_{d}v_{d})}{\partial \xi}+\lambda \epsilon \frac{\partial}{\partial \eta}(n_{d}v_{d})=0
\label{eq:continuity_stretched}\\
&-&\frac{\partial v_{d}}{\partial \xi}+\frac{v_{d}}{\lambda}\frac{\partial v_{d}}{\partial \xi}+v_{d}\epsilon\frac{\partial v_{d}}{\partial \eta}-\frac{Z_{d} e}{m_{d}\lambda}\frac{\partial\phi}{\partial\xi}=\epsilon \frac{Z_{d} e}{m_{d}}\frac{\partial \phi}{\partial\eta}
\label{eq:momentum_stretched}\\
&\epsilon&\frac{\partial^{2}\phi}{\partial\xi^{2}}+\lambda^{2}\epsilon^{3}\frac{\partial^{2}\phi}{\partial\eta^{2}}+2\lambda\epsilon^{2}\frac{\partial^{2}\phi}{\partial\xi\partial\eta}-\epsilon^{2}\frac{\partial\lambda}{\partial\eta}\frac{\partial\phi}{\partial\xi}+4\pi e\lambda^{2}\nonumber\\
&\times&\left[n_{i0}\text{exp}\left( \frac{-e\phi}{T_{i}}\right) -n_{e0}\text{exp}\left( \frac{e\phi}{T_{e}}\right) -Z_{d}n_{d} \right]=0
\label{eq:Poisson_stretched}
\end{eqnarray}
%%%%%%%%%%%%%%%%%%%%%%%%%%%%%%%%%%%
We now substitute Eqs.~\ref{eq:Expansion} into  Eqs.~\ref{eq:continuity_stretched}~--~\ref{eq:Poisson_stretched} and set the sum of  the terms of the same power of $\epsilon$ to zero in each of these equations. \\
To first order in $\epsilon$  we then get,
%%%%%%%%%%%%%%%%%%%%%%%%%%%%%%%%%%%
\begin{equation}
-\lambda\frac{\partial n_{d1}}{\partial \xi}+n_{d0}\frac{\partial v_{d1}}{\partial \xi}+v_{d0}\frac{\partial n_{d1}}{\partial \xi}+\lambda n_{d0}\frac{\partial v_{d0}}{\partial \eta}+\lambda v_{d0}\frac{\partial n_{d0}}{\partial \eta}=0
\label{eq:first_continuity}
\end{equation}
%%%%%%%%%%%%%%%%%%%%%%%%%%%%%%%%%%%
%%%%%%%%%%%%%%%%%%%%%%%%%%%%%%%%%%%
\begin{equation}
-\lambda\frac{\partial v_{d1}}{\partial \xi}+v_{d0}\frac{\partial v_{d1}}{\partial \xi}+\lambda v_{d0}\frac{\partial v_{d0}}{\partial \eta}-\frac{Z_{d} e}{m_{d}}\frac{\partial\phi_{1}}{\partial \xi}-\frac{Z_{d} e}{m_{d}}\lambda\frac{\partial\phi_{0}}{\partial \eta}=0
\label{eq:first_momentum}
\end{equation}
%%%%%%%%%%%%%%%%%%%%%%%%%%%%%%%%%%%
%%%%%%%%%%%%%%%%%%%%%%%%%%%%%%%%%%%
\begin{equation}
4 \pi e \lambda^{2} Z_{d} n_{d1}+4 \pi e^{2} \lambda^{2}  \frac{n_{e0}}{T_{e}}\phi_{1}+4 \pi e^{2} \lambda^{2}  \frac{n_{i0}}{T_{i}} \phi_{1}=0
\label{eq:first_Poisson}
\end{equation}
%%%%%%%%%%%%%%%%%%%%%%%%%%%%%%%%%%%
Eq.~\ref{eq:first_Poisson} can be expressed as:
%%%%%%%%%%%%%%%%%%%%%%%%%%%%%%%%%%%
\begin{equation}
n_{d1}=\frac{-e\left(n_{i0}T_{e}+n_{e0}T_{i} \right) }{Z_{d}T_{e}T_{i}}\phi_{1}
\label{eq:relation_nd1_phi1}
\end{equation}
%%%%%%%%%%%%%%%%%%%%%%%%%%%%%%%%%
Integrating Eq.~\ref{eq:first_continuity} with respect to $\xi$ to get the relation between $n_{d1}$ and $v_{d1}$ as:
\begin{equation}
v_{d1}=\frac{(\lambda-v_{d0})}{n_{d0}}n_{d1}-\lambda \xi\frac{\partial v_{d0}}{\partial \eta}-\lambda \xi\frac{v_{d0}}{n_{d0}}\frac{\partial n_{d0}}{\partial \eta} 
\label{eq:relation_nd1_vd1}
\end{equation}
%%%%%%%%%%%%%%%%%%%%%%%%%%%%%%%%%%%
Eqs.~\ref{eq:first_continuity}--\ref{eq:first_momentum} are then integrated with respect to $\xi$ with the boundary conditions  $n_{d0} \rightarrow 1$ and ($ \phi_0, v_{d0}, n_{d1}, v_{d1},\phi_{1} \rightarrow 0$) as $\xi \rightarrow \infty$. We then eliminate $n_{d1} $ and $v_{d1} $, to obtain the following expression for $\phi_{1}$, 
%%%%%%%%%%%%%%%%%%%%%%%%%%%%%%%%%%%
%%%%%%%%%%%%%%%%%%%%%%%%%%%%%%%%%%%
\begin{equation}
\phi_{1}=\frac{m_{d}n_{d0}C_{DA}^{2}\lambda\xi}{Z_{d} e}\left[\frac{\lambda\frac{\partial v_{d0}}{\partial\eta}+(\lambda-v_{d0})\frac{v_{d0}}{n_{d0}}\frac{\partial n_{d0}}{\partial\eta}-\frac{Z_{d} e}{m_{d}}\frac{\partial \phi_{0}}{\partial\eta}}{C_{DA}^{2}-(\lambda-v_{d0})^{2}} \right]. 
\label{eq:expression_for_phi1}
\end{equation}
%%%%%%%%%%%%%%%%%%%%%%%%%%%%%%%%%%%
where, $ C_{DA}=\sqrt{\frac{Z_{d}^{2}T_{e}T_{i}}{m_{d}(n_{i0}T_{e}+n_{e0}T_{i})}} $ is the dust-acoustic speed. The right-hand side of Eq.~\ref{eq:expression_for_phi1} contains only zeroth order quantities, whereas $\phi_{1}$ is a first order variable. To obtain a finite value of $\phi_{1}$ one requires to separately set the numerator and denominator on the R.H.S. to zero \cite{Singh_1998}. Setting the numerator to zero gives us, 
%%%%%%%%%%%%%%%%%%%%%%%%%%%%%%%%%%%
\begin{equation}
\lambda\frac{\partial v_{d0}}{\partial\eta}+(\lambda-v_{d0})\frac{v_{d0}}{n_{d0}}\frac{\partial n_{d0}}{\partial\eta}-\frac{Z_{d} e}{m_{d}}\frac{\partial \phi_{0}}{\partial\eta} =0
\label{eq:numerator}
\end{equation}
which is a self-consistent relation between the gradients of the equilibrium quantities $n_{d0}, v_{d0}$ and $\phi_{0}$. By setting the denominator to zero we get,
\begin{equation}
\lambda=C_{DA}+v_{d0}
\label{eq:denominator}
\end{equation}
which defines the velocity of the wave frame.
%%%%%%%%%%%%%%%%%%%%%%%%%%%%%%%%%%%
Equating the coefficients of $ \epsilon^{2} $ in Eqs.~\ref{eq:continuity_stretched}--\ref{eq:Poisson_stretched} to zero, the following set of equations are obtained:
%%%%%%%%%%%%%%%%%%%%%%%%%%%%%%%%%%%
\begin{eqnarray}
-(\lambda-v_{d0})\frac{\partial n_{d2}}{\partial\xi}&+&n_{d0}\frac{\partial v_{d2}}{\partial\xi}+\frac{\partial}{\partial\xi}(n_{d1}v_{d1})+\nonumber\\
&\lambda&\frac{\partial}{\partial\eta}(n_{d0}v_{d1}+n_{d1}v_{d0})=0
\label{eq:second_continuity}\\
-(\lambda-v_{d0})\frac{\partial v_{d2}}{\partial\xi}&-&\frac{Z_{d} e}{m_{d}}\frac{\partial \phi_{2}}{\partial\xi}+v_{d1}\frac{\partial v_{d1}}{\partial\xi}+\lambda v_{d0}\frac{\partial v_{d1}}{\partial\eta}+\nonumber\\
&\lambda&v_{d1}\frac{\partial v_{d0}}{\partial\eta}-\frac{Z_{d} e}{m_{d}}\lambda\frac{\partial \phi_{1}}{\partial\eta}=0
\label{eq:second_momentum}\\
\frac{\partial^{2}\phi_{1}}{\partial\xi^{2}}+4\pi e\lambda^{2}&\times&\left[\frac{-e(n_{i0}T_{e}+n_{e0}T_{i})}{T_{e}T_{i}}\phi_{2}+\right. \left(\frac{n_{i0}}{T_{i}^{2}}-\frac{n_{e0}}{T_{e}^{2}} \right) \nonumber\\
&&\left. \times \frac{e^{2}\phi_{1}^{2}}{2}-Z_{d}n_{d2} \right]=0 
\label{eq:second_Poisson}
\end{eqnarray}
%%%%%%%%%%%%%%%%%%%%%%%%%%%%%%%%%%%
%It is found in the experiments that the equilibrium dust density remains almost constant along z-axis. In addition, it is also evident from the experiments that the equilibrium velocity of the dust component becomes negligible. These experimental findings essentially justify the conditions  $\frac{\partial n_{d0}}{\partial\eta}=0$ and $v_{d0} << \lambda$.
Taking the derivative of Eq.~\ref{eq:second_Poisson} with respect to $ \xi $ to get
\begin{eqnarray}
 \frac{\partial^{3}\phi_{1}}{\partial\xi^{3}}+4\pi e\lambda^{2}&\times&\left[\frac{-e(n_{i0}T_{e}+n_{e0}T_{i})}{T_{e}T_{i}}\frac{\partial\phi_{2}}{\partial \xi}+\right. \left(\frac{n_{i0}}{T_{i}^{2}}-\frac{n_{e0}}{T_{e}^{2}} \right) \nonumber\\
&&\left. \times e^{2}\phi_{1}\frac{\partial\phi_{1}}{\partial\xi}-Z_{d}\frac{\partial n_{d2}}{\partial\xi} \right]=0 
\label{eq:Poisson_derivative}
\end{eqnarray}
Now, substituting the second order quantities $\frac{\partial n_{d2}}{\partial\xi}$ from Eq.~\ref{eq:second_continuity} and $\frac{\partial\phi_{2}}{\partial \xi}$ from Eq.~\ref{eq:second_momentum} in Eq.~\ref{eq:Poisson_derivative}, we obtain
%%%%%%%%%%%%%%%%%%%%%%%
\begin{eqnarray}
\hspace*{-0.1in}\frac{\partial^{3}\phi_{1}}{\partial\xi^{3}}&+&4\pi e\lambda^{2}\left[\left(\frac{n_{i0}}{T_{i}^{2}}-\frac{n_{e0}}{T_{e}^{2}} \right) e^{2}\phi_{1}\frac{\partial\phi_{1}}{\partial\xi} +\frac{m_d(n_{i0}T_{e}+n_{e0}T_{i})}{Z_dT_{e}T_{i}}\right.\nonumber\\
\label{eq:eliminating_second}
&&\hspace*{-0.25in}\times \left.\left(-v_{d1}\frac{\partial v_{d1}}{\partial\xi}-\lambda\frac{\partial(v_{d0}v_{d1})}{\partial\eta}+\lambda\frac{Z_{d}e}{m_{d}}\frac{\partial\phi_{1}}{\partial\eta} \right)-\right. \\
&&\hspace*{-0.3in}\left.\frac{Z_{d}}{\lambda-v_{d0}}\left(\frac{\partial (n_{d1}v_{d1})}{\partial\xi}+\lambda\frac{\partial }{\partial\eta}(n_{d0}v_{d1}+n_{d1}v_{d0}) \right) \right]=0 \nonumber
\end{eqnarray}
%%%%%%%%%%%%%%%%%%%%%%%
We then substitute $\phi_{1}$ from Eq.~\ref{eq:relation_nd1_phi1} and $v_{d1} $ from Eq.~\ref{eq:relation_nd1_vd1} in terms of $n_{d1}$ in Eq.~\ref{eq:eliminating_second}. In addition, we also put the values of $\frac{\partial v_{d0}}{\partial\eta}$ and $v_{d0}$ from Eq.~\ref{eq:numerator} and Eq.~\ref{eq:denominator} in Eq.~\ref{eq:eliminating_second} to obtain the following form of the modified-KdV equation:
%%%%%%%%%%%%%%%%%%%%%%%%%%%%%%%%%%%
\begin{equation}
\frac{\partial n_{d1}}{\partial\eta}+P\frac{\partial^{3}n_{d1}}{\partial\xi^{3}}+Qn_{d1}\frac{\partial n_{d1}}{\partial\xi}-R\frac{\partial\phi_{0}}{\partial\eta}n_{d1}-S\frac{\partial n_{d0}}{\partial\eta}n_{d1}=0.
\label{eq:mKdV}
\end{equation}
%%%%%%%%%%%%%%%%%%%%%%%%%%%%%%%%%%%
where the coefficients $P$, $Q$, $R$ and $S$  are given by, \\
\begin{eqnarray}
P&=&\frac{1}{2\lambda^{3}(1+\sigma_{ie})}\nonumber\\
Q&=&\frac{[(1-\sigma_{ie}^{2}) Z_{d}-2(1+\sigma_{ie})^{2}]}{(1+\sigma_{ie})^{2}\lambda(1+Z_{d})}\nonumber\\
R&=&\frac{Z_{d} }{\lambda^{2}(1+Z_{d})}\nonumber\\
S&=&\frac{(1+\sigma_{ie})\lambda^{2}}{ Z_{d}}.\nonumber
\end{eqnarray}
%%%%%%%%%%%%%%%%%%% FIGURE   %%%%%%%%%%%%%%%%%%%%%
 \begin{figure}[h]
\includegraphics[scale=0.65]{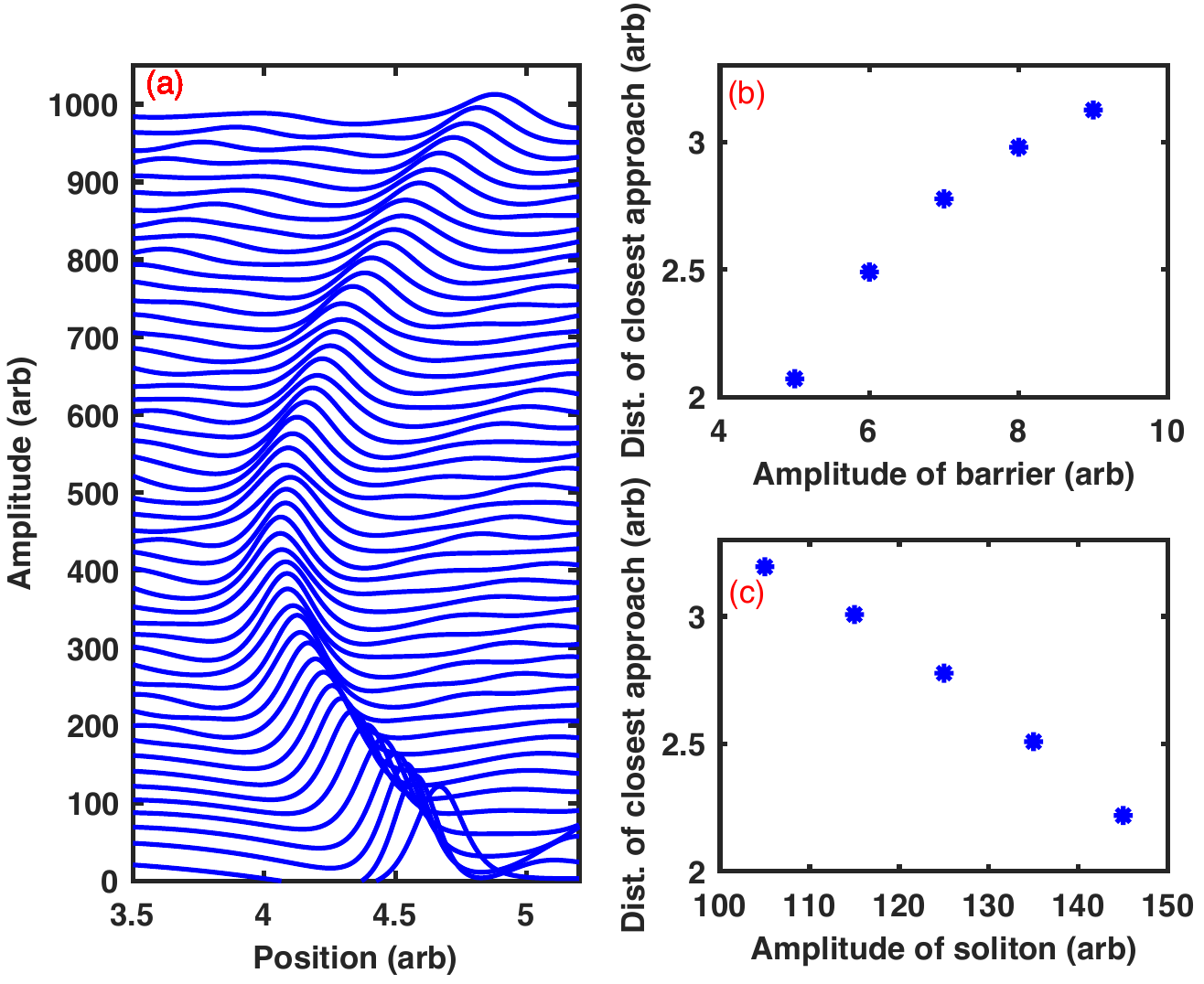}
\caption{\label{fig:fig10} (a) Time evolution of soliton solution. Variation of the distance of the closest approach with (b) the strength of potential and (c) the amplitude of the solitary wave. }
\end{figure}  
%%%%%%%%%%%%%%%%%%%%%%%%%%%%%%%%% 
Where, $\sigma_{ie}$ is the ratio of ion temperature to electron temperature. It is to be noted that all the variables used in the modified KdV equation (Eq.~\ref{eq:mKdV}) as well as in its coefficients are in normalized units, where  space is normalized by $\sqrt{\frac{T_{i}}{4\pi Z_{d}n_{d0}e^{2}}} $, time is normalized by $\sqrt{\frac{m_{d}}{4\pi n_{d0}Z_{d}^{2}e^{2}}} $, velocity is normalized by $\sqrt{\frac{Z_{d}^{2}T_{i}}{m_{d}}} $,  density, $ n_{s} $ is normalized by $ n_{s0} $ and electrostatic potential, $ \phi $ is normalized by $ \frac{T_{i}}{e} $. The fourth term in Eq.~\ref{eq:mKdV} arises due the presence of the external space varying potential which replicates the sheath potential around the wire akin to the experimental situation. \par
For a qualitative understanding of the experimental results we have solved the above modified KdV equation numerically. The sheath potential around the biased wire has an approximate Gaussian shape in space as discussed in Section  \ref{sheath} and consequently the equilibrium density also acquires a similar form in the region of the sheath.  Our experimental measurements show that the density variation is very small compared to the potential variation and we therefore neglect the contribution proportional to $S$ in Eqn.~\ref{eq:mKdV} for our numerical computations. Taking $\phi_{0} \approx \left(\text{a exp}[\frac{-(\eta-b)^{2}}{c}]\right)$ we display our numerical results of the evolution of the solitary pulse along the axial direction in Fig.~\ref{fig:fig10}(a)  for $a=7$, $b=5$ and $c=30$. \par
 As can be seen the simulated profiles of perturbed densities show a behaviour that is qualitatively  similar to the experimental results. As found in the experiments, the soliton initially propagates in the forward direction and then it reflects due to the presence of the external potential. The point of the closest approach mainly depends on the strength of external potential $(a)$ and the initial amplitude (or velocity) of the excited solitons. The variation of distance of closest approach with the strength of potential (barrier) is shown in fig.~\ref{fig:fig10}(b). Similar to the experimental observation, the distance of the closest approach increases with the increase of the amplitude of the Gaussian function ($a$) used in the simulation. Furthermore, the variation of the distance of closest approach with the amplitude of the soliton is also investigated and is shown in fig.~\ref{fig:fig10}(c). As expected, the solitons with higher amplitudes penetrate deeper into the external potential. Thus, the above KdV model provides a good qualitative description of the present set of experimental results. 
%%%%%%%%%%%%%%%%%%%%%%  
\section{Conclusion}\label{sec: summary}
To conclude, an experimental demonstration of reflection of a dust acoustic solitary wave by a negative sheath potential is presented for the first time in a dusty plasma medium. The experiments are performed in the Dusty Plasma Experimental (DPEx) device in which dusty plasma is created in a DC Ar plasma environment. The dust acoustic solitary wave is excited by modulating the plasma by applying a negative Gaussian short pulse over the discharge voltage. The perturbed dust density wave is characterized as a solitary wave by measuring its amplitude ($A$) and width ($L$) over time. It is found that the soliton parameter ($AL^2$) maintains a constant value like a KdV type soliton in the passage of its journey after it grows fully.  The solitary wave moves towards the potential barrier created by the sheath around a biased copper wire and gets reflected after interacting with the sheath potential. The interaction of the solitary wave with the potential barrier is investigated in detail by altering the strength of the sheath potential and the initial amplitudes of the solitary wave. It is found that the distance of the closest approach of the solitary wave becomes more for the higher strength of the potential barrier and the lower amplitude of the solitary wave. An emissive probe is employed to estimate the sheath thickness and the sheath potential around the copper wire. It is found that the sheath thickness (and strength of the potential barrier) decreases when the floating wire is connected with a resistor with higher values. To provide a theoretical understanding of our experimental findings, we have developed a model equation in the form of a modified KdV equation.  The numerical solutions of this KdV equation show a good qualitative agreement with our main experimental findings, namely, the dependence of the magnitude of closest approach of the solitary wave on the soliton wave amplitude and the amplitude of the barrier potential.  Our findings should be useful in further experimental and theoretical studies of such phenomena in a laboratory dusty plasma as well as in space plasma environments. 
 \begin{acknowledgments}
A.S. is thankful to the Indian National Science Academy (INSA) for their support under the INSA Senior Scientist Fellowship scheme. K.K. would like to acknowledge Minsha Shah for her help in designing the wave exciter circuit. \\ \\
\noindent \textbf{DATA AVAILABILITY}\\\\
The data that support the findings of this study are available from the corresponding author upon reasonable request.\\\\
\end{acknowledgments}
\noindent \textbf{References}
\end{document}